\documentclass[11pt,twoside]{article}
\usepackage{asp2004}
\usepackage{psfig}
\usepackage{epsf}
\usepackage{graphics}
\usepackage{lscape}
\def\edcomment#1{\iffalse\marginpar{\raggedright\sl#1\/}\else\relax\fi}
\reversemarginpar

\begin{document}
\title{A High Resolution HI Study of Selected Virgo Galaxies\\
{\small - Preliminary Results on Gas Morphology \& Extra-Planar Gas}}
\author{Aeree Chung and J. H. van Gorkom}
\affil{Department of Astronomy, Columbia University, 
550 West 120th Street, New York, NY 10027, USA}
\author{Jeffery D. P. Kenney}
\affil{Yale University Astronomy Department, 
P. O. Box 208101, New Haven, CT 06520-8101, USA}
\author{Bernd Vollmer}
\affil{CDS, Observatoire astronomique de Strasbourg, UMR 7550, 
11 rue de I'universit$\acute{e}$, 67000 Strasbourg, France}


\begin{abstract}
We present preliminary results of VLA HI imaging of selected Virgo cluster 
galaxies. The goal is to study environmental effects on galaxy evolution. 
Our sample of 41 galaxies is spread throughout the cluster and spans a wide 
range in star formation properties. Here we present the total HI maps of 
13 galaxies. We find a number of galaxies with extended HI tails, 
almost all pointing away from the cluster center. Truncated HI disks are found 
close to the center but also in the outer region. Some galaxies near the
cluster center show compression of the gas on one side. Multiwavelength
data of NGC~4569 and kinematics on 
NGC~4396 indicate that some of the HI is extra-planar. 
These preliminary results on the HI morphology already suggest 
that a variety of environmental effects such as ICM-ISM interactions, 
harassment, tidal interactions or mergers may be at work to 
affect the evolution of galaxies.
\end{abstract}
\thispagestyle{plain}

\section{VLA HI STUDY OF SELECTED VIRGO GALAXIES}
It has long been known that galaxies in the dense cluster environments 
systematically differ from those in the field in their morphology, stellar 
populations, gas fractions and gas distributions. More interestingly,
observations of galaxy clusters at intermediate redshift show that these
properties change with redshift. This could indicate that the dense cluster 
environment speeds up the evolution of galaxies. Various mechanisms affecting
the evolution of galaxies have been suggested such as ram-pressure stripping, 
merging, tidal interaction, harassment or starvation. In spite of the 
abundance of statistical 
studies on clusters at intermediate redshifts \citep[e.g.][]{detal99,petal99}
there is a lack of indepth studies of
individual galaxies which will further constrain the environmental effects 
in clusters. Our goal is to do a detailed study of galaxies that are 
currently being affected by the cluster environment.
Virgo is ideal for this 
purpose. Its nearness allows us to study details, and as a dynamically young 
cluster it shows a variety of processes at work to affect the galaxies.
Different mechanisms can be traced by HI observations.
Since the first Virgo survey in early 1980s the sensitivity of the VLA 
(Very Large Array) has been significantly improved by almost a factor 
of 10 with a comparable resolution. As several Virgo galaxies observed 
with a higher sensitisity show \citep[e.g.][]{pvm93,pm95,vcbbd99}, the 
newer HI data will provide much more detailed (and possibly never seen)
structures.

\begin{figure}[!t]
\caption{Spatial distributions of the sample galaxies are shown overlaid 
on ROSAT x-ray map (left) and the galaxy number density map (right) of Virgo. 
The galaxies observed in this year are drawn with ellipses which are 5 times 
bigger than the actual size and the rest galaxies in the complete sample are 
indicated in crosses.}
\end{figure}

\section{SAMPLE SELECTION and OBSERVATIONS}
We selected 41 spirals (S0/a$-$Sm) in the Virgo cluster (Figure 1). 
These galaxies cover a factor of 50 in mass of the cluster and span
a wide range in star formation properties \citep{kk04}.
They are located throughout the cluster, from the dense region close 
to the center to the low density outer parts. Compared to previous surveys 
\citep[e.g.][]{w88,cvbk90} with brightness cutoff of 
$10-11$ in $B_T$, our sample also contains fainter systems of 
$12-13$ mag which are more likely to be affected by stripping or
gravitational interactions due to their low masses.  

The observations of the 13 galaxies were made between February 
and May 2004 in VLA C-short array. We integrated $\approx$8 hours on each 
source with a total bandwidth of 3.25~MHz and a channel separation of 
24.4~kHz (5.2 km s$^{-1}$ at 21~cm). Online hanning smoothing 
has been applied and the resulting velocity 
resolution is 10.4 km s$^{-1}$. We typically reached 
$\rm\approx$0.3~mJy per beam (typically $16.5''\times16.5''$) per channel 
(10.4 km s$^{-1}$). This corresponds to a 3$\bf\sigma$ surface density 
sensitivity of 4$\rm\times$10$^{19}$~cm$^{-2}$. 
In some cases (e.g. NGC~4294, NGC~4299, NGC~4383 and NGC~4694) the cubes
were spatially convolved in order to bring out faint structures.

\section{TOTAL HI INTENSITY MAPS}

\begin{figure}[!t]
\caption{The total HI maps of 13 galaxies. Solid ellipses indicate
the optical size ($D_{25}$, ellipticity and position angle in $B$).
The distribution of the galaxies are shown overlaid on ROSAT X-ray 
map and the galaxy density map with numbers which are assigned in
order of NGC number. The beam size is indicated on the left bottom
corner of each box. The lowest contour level is 
$\approx0.8-1.6\times10^{19}$ cm$^{-2}$ beam$^{-1}$ assuming 
$\sim16.5''\times16.5''$, the mean size of the beam of our 
observations.} 
\end{figure}

In Figure 2 we present the total HI maps of 13 galaxies. NGC~4351,
NGC~4396, and NGC~4189 decline more sharply in HI surface brightness 
on the side toward the cluster center. These galaxies are at 
intermediate distances from M87 
(1.7$-$4.3 deg) and are likely to be experiencing on-going ICM-ISM
pressure.

Extended tails are seen in a number of galaxies. NGC~4294/9, 
NGC~4351, NGC~4396, NGC~4424, NGC4651, and NGC4698 are found at
a range of distances ($2.4-5.8$ deg from M87). No obvious companions
are found around these galaxies and none of their tails except
possibly NGC~4424 and NGC 4651 seems to be related to tidal interactions.
Especially 
we note that NGC 4294 and NGC 4299 which are only 0.1 deg apart from 
each other and at similar redshifts, show tails in the same direction,
unlike what happens in tidal interactions.
Even though we do not have a coherent three dimensional picture of Virgo, 
it is worth mentioning that all the tails except for one case 
(NGC~4651) are pointing away from the cluster center. 

The southern extension in HI of NGC~4424 could be related to 
the giant elliptical NGC~4472 which is located 1.56 deg away to south,
either through a tidal interaction or an ICM-ISM interaction \citep{bjfke04}.
NGC 4651 is a peculiar case in a sense that its optical tail 
and the gas tail are extended in opposite directions, to east for the stellar
extension and to west for the gas tail, suggesting a minor merger.
Gas accretion or tidal interaction is also found in some galaxies such 
as NGC~4383 (a small gas blob in SE) and NGC~4694 (tidally interacting 
with a low surface brightness system VCC~2062).

There are several galaxies with 
truncated HI disks such as NGC~4064, NGC 4424, 
NGC~4569, and NGC~4580. NGC~4424 and NGC~4569 are located in high density 
regions in a sub-cluster or in the center of the cluster (NGC~4569 is
discussed further in the following section) and ICM-ISM pressure likely 
has caused the truncation in the gas. However NGC~4064 and NGC~4580 are 
exceptional in a sense that both of them are located in low density 
environments with projected distances of 8.8 and 7.2 
degrees from M87 \citep[3.1 and 2.5 virial radii;][]{ts84},
respectively. The most likely explanation is that they have gone through 
the center and are on their way out. Even if galaxies have not gone through
the the highest density regions,
ICM-ISM interactions could still happen at further distances from the cluster
center when  the galaxy interacts with locally enhanced ICM due to sub 
cluster-cluster 
merging  as suggested by \citet{kvv04} for
NGC 4522. 

Deep optical and H$\alpha$ images have already been taken.
More recently we have been granted GALEX (Galaxy Evolution Explorer) time 
for the entire sample of 41 galaxies. The UV data will allow us to trace
the timescales on processes at work. Combined by multiwavelength data
and also compared with simulations eventually we will get more clear 
understandings of galaxy evolutions in the cluster environments.

\section{EXTRA-PLANAR GAS IN VIRGO CLUSTER GALAXIES}

NGC~4569 is one of the closest galaxies to the cluster center in our
sample and it is severely deficient in HI ($\sim10$\% of the normal 
HI for its size and type). Both the H$\alpha$ and HI are truncated, 
and only extend to 30\% of its stellar disk. An anomalous arm is seen to 
the west both in HI and H$\alpha$ (Figure 3) and the HI arm seems
to be interrupted at the location of the H$\alpha$ starburst outflow 
nebulosity. The HI arm extends for 3 arcmin($\approx$13 kpc) while the
H$\alpha$ nebulosity near the minor axis on the NW side extends up to 
6 kpc from the nucleus. Some H$\alpha$ filaments are also observed on the SE
side, but only within $5''$ of the nucleus, and there is no large-scale
H$\alpha$ nebulosity in the SE like that in the NW.
These diffuse structures seen along the minor axis must be 
a product of a nuclear starburst and have been disrupted by ICM pressure 
(nebulosity as the superbubble). 
The ICM pressure must be stronger on 
the SE to see that the nebulosity is weaker on this side and the more 
extended HI structure on the other side seems to be consistent with 
this picture. This arm thus must be extra-planar, because it appears to be
interacting with the minor axis starburst outflow, and because it lacks 
significant dust extinction, suggesting that it lies behind the disk.
The arm resembles the features that are seen in some 
phases of ICM-ISM interaction simulations \citep{vcbd01,ss01}
It has been suggested that the 
stripped gas can form one extra-planar arm by the combination of wind 
pressure and galaxy rotation \citep[see also ][]{vbcvh04}.

\begin{figure}[!t]
\caption{NGC~4569. From left to right, HI contours, hard x-ray,
and soft x-ray overlaid on H$\alpha$ \citep{ketal04}. The contour 
levels of the HI map are 1.77, 5.3, 10.6, 17.7, 26.6 and 37.2$\times10^{19}$
cm$^{-2}$ beam$^{-1}$.}
\end{figure}

NGC~4396 also show signatures of extra-planar gas. 
The kinematics of this gas are especially interesting (Figure 4). 
Note how the velocities in the component north of the disk are all
redshifted w.r.t. nearby disk on both sides of minor axis suggesting
non-circular motions, possibly indication of the presence of extra-planar 
gas. Simulations, estimating ICM pressure in this region or studying 
stellar population will be helpful for sorting out the interaction 
parameters and history. Some other cases of extra-planar gas in 
Virgo cluster galaxies have been studied in highly inclined systems. 
There are several pieces of evidence that NGC 4522 is experiencing ongoing
ICM-ISM stripping despite its location far (but similar to
the distance of NGC 4396) from the cluster center \citep{kvv04,vbkv04}.
See also the case of NGC 4402 by Crowl et al. (2004).

\begin{figure}[!t]
\caption{Position velocity diagrams of NGC~4396 along the major/minor
axis at different scale heights/radii. The contour levels of the HI 
map are (1.25+5n)$\times10^{19}$ cm$^{-2}$ beam$^{-1}$ (n=0, 1, 2, 3, ...).
The thick black line in the velocity field represents -128 km s$^{-1}$,
the systemic velocity. Thin black and white lines indicate
lower and higher velocity sides respectively with $\Delta{V}=10$ km s$^{-1}$.}
\end{figure}

\acknowledgments{
This work was supported by NSF grants to Columbia and Yale University. 
The National Radio Astronomy Observatory is a facility of the National 
Science Foundation operated under cooperative agreement by Associated 
Universities, Inc.}

\end{document}